\documentclass[aps,floats,epsf, twocolumn]{revtex4}
 \usepackage{graphics}

\begin{document}

\title{Conserved charges of order-parameter textures in  Dirac systems}
\author{Igor F. Herbut$^1$, Chi-Ken Lu$^1$, and Bitan Roy$^2$}

\affiliation{$^1$Department of Physics, Simon Fraser University, Burnaby, British Columbia, Canada V5A 1S6\\
$^2$ National High Magnetic Field Laboratory, Florida State University, Tallahassee, Florida 32306, USA}

\begin{abstract}
A simple expression for the induced fermion current in the presence of a texture in mass-order-parameters in two-dimensional condensed-matter Dirac systems is derived using  the representation theory of Clifford algebras. In particular, it is shown that every texture in three mutually anticommuting order parameters, in graphene for example, implies an induced density of a properly defined conserved charge. The sufficient condition for the general charge to be the familiar electrical charge is that the remaining two anticommuting order parameters allowed by the particle-hole symmetry are the two phase components of some superconducting order. This allows eight different types of electrically charged textures in graphene or in the $\pi$-flux Hamiltonian on the square lattice. Generalized charge of mass-textures on the surfaces of thin films of topological insulators, or in spinless Dirac fermions hopping on the honeycomb lattice is also discussed.
 \end{abstract}
\maketitle

\vspace{10pt}

\section{Introduction}

    It is well known that in Fermi systems described by the  minimal Dirac Hamiltonian in two spatial dimensions with a general and inhomogeneous mass term,
\begin{equation}
H_0 (\vec{m}) = i \gamma_0 ( \hat{p}_1 \gamma_1 + \hat{p}_2  \gamma_2 +i m_1 + m_2 \gamma_3 + m_3 \gamma_5),
\end{equation}
where $m_i = m_i (\vec{x})$, the induced fermionic current is proportional to the conserved topological mass-current: \cite{goldstone,jaroszewicz, abanov}
\begin{equation}
\langle \bar{\Psi} \gamma_\mu \Psi \rangle = \frac{1}{8\pi} \epsilon_{\mu \nu \lambda} \vec{m} \cdot (\partial_\nu \vec{m} \times \partial_\lambda  \vec{m}).
\end{equation}
Here the Hermitian four-dimensional $\gamma$-matrices satisfy the anticommutation relations $\{ \gamma_\mu, \gamma_\nu \} = 2 \delta_{\mu \nu}$, and $\bar{\Psi}= \Psi^\dagger \gamma_0$. This, for example, implies that a Skyrmion configuration in the mass $\vec{m}$ in total carries a single unit of the $U(1)$ gauge charge.  Indeed, in graphene, as a condensed matter embodiment of the Dirac equation, the Skyrmion in the vector order parameter that describes the anomalous spin Hall state can be understood as two identical copies of the above Dirac Hamiltonian, yielding the total electrical charge of two.\cite{grover,lu} Bilayer graphene, the absence of Lorentz invariance of its Hamiltonian notwithstanding, can also be shown to support  Skyrmions in the anomalous spin Hall state, but of the twice larger charge of four.\cite{lu, moon} This purely topological  mechanism of electron pairing opens up new and unconventional possibilities for the emergence of superconductivity in such Dirac systems \cite{pasha, dhlee}, and has also been discussed in the context of deconfined criticality. \cite{grover, moon, tanaka}

  In the single and double layer graphene, however, the low-energy Dirac (or more precisely, Dirac-like, for a bilayer) Hamiltonian is eight-dimensional, and if Cooper pairing is to be accommodated, sixteen-dimensional matrix operator. It therefore may not come as a surprise that when these Hamiltonians can be block-diagonalized,  with the elementary blocks of the form in Eq. (1), the electrical charges of the individual blocks may mutually cancel. This is precisely what happens if the Skyrmion is formed out of the three components of the N\' eel antiferromagnetic order parameter in the single-layer, \cite{tanaka} or out of the layer-antiferromagnetic order parameter in the  bilayer graphene, \cite{lu} when the residual chiral symmetry of the Hamiltonian guarantees vanishing of the electrical charge. In this sense the anomalous spin Hall state may appear as if enjoying a privileged position among many possible order parameters in the Dirac electronic systems, and being possibly the unique ``dual" to superconductivity. \cite{herbut3}

This raises the obvious question addressed in this paper: what exactly is the condition on the three mass terms (order parameters) being wound, so that the Skyrmion, or a more general texture, develops the {\it electrical} charge? The answer follows from a more general result, which we discuss first, and prove later in the text: if the low-energy excitations in presence of a texture in three mutually anticommuting order parameters $m_i$, $i=1,2,3$ are described by the general Dirac-Nambu particle-hole-symmetric Hamiltonian
\begin{equation}
H= i\Gamma_0 (\hat{p}_1\Gamma_1 + \hat{p}_2 \Gamma_2 +i  m_1 + m_2 \Gamma_3 + m_3\Gamma_5),
\end{equation}
where the Hermitian $\Gamma$-matrices are {\it d-dimensional} ($d=8,16$), and $\{ \Gamma_\mu, \Gamma_\nu \} = 2 \delta_{\mu \nu}$, then the induced fermion current is given by the following expression:
\begin{equation}
\langle \bar{\Psi} \Gamma_\mu X \Psi \rangle = \frac{d}{64\pi} \epsilon_{\mu \nu \lambda} \vec{m} \cdot (\partial_\nu \vec{m} \times \partial_\lambda  \vec{m}),
\end{equation}
with $\bar{\Psi} = \Psi^\dagger \Gamma_0$, and with the matrix $X$ as
\begin{equation}
X = \Gamma_0 \Gamma_1 \Gamma_2 \Gamma_3 \Gamma_5.
\end{equation}
This means that {\it every} texture in three mutually anticommuting order parameters, insulating or superconducting, carries a distribution of some generalized conserved ``charge" $\langle \Psi^\dagger X \Psi \rangle$,  determined by the matrix $X$, which also commutes with the Hamiltonian $H$.  The particle-hole symmetry prohibits $X$ to  simply be the  unit matrix, but the general charge reduces to the electrical charge if the matrix $X$ happens to be the operator for the particle number in the given representation.

In graphene, as a representative of a class of two-dimensional condensed matter Dirac systems which obey parity and the time reversal symmetries for each spin component separately,\cite{herbut2} there are two sublattices, two Dirac points (or valleys), and two projections of the electron spin, so including the  particle-hole doubling, it follows that the relevant dimension of the Dirac-Nambu Hamiltonian is $d=16$. Quite generally for this class of systems, the particle-hole symmetry allows 36 different mass terms, which can be grouped into 56 fivetuplets of mutually anticommuting order parameters.\cite{ryu, remark0} Let us consider the particular, and as it turns out the unique, (Appendix C) fivetuplet to which the matrices $\Gamma_i$, $i=0,3,5$
belong. One can always choose the particularly convenient  ``Majorana" representation in which the matrices $\Gamma_1$ and $\Gamma_2$ are real, and $\Gamma_i$, $i=0,3,5,6,7$ are imaginary, where $\Gamma_6$ and $\Gamma_7$ are the further two matrices that complement $\Gamma_0,\Gamma_3,\Gamma_5$ to the aforementioned fivetuplet. \cite{ryu, herbut1} In this representation the Hamiltonian $H$ is imaginary, and the (antilinear) particle-hole symmetry, under which $H$ changes sign, is just the complex conjugation. It will then follow that
   \begin{equation}
   X= i \Gamma_6 \Gamma_7,
   \end{equation}
i. e. the matrix $X$ is the generator of ``rotations" between the remaining two matrices in the fivetuplet, $\Gamma_6$ and $\Gamma_7$. If $X$ is the number operator, the latter two must stand for the two phase components of a superconducting order parameter. Since in graphene-like systems there are four possible types of the superconducting mass-gaps, \cite{roy} and each pair of anticommuting mass-matrices belongs to two different fivetuplets, \cite{herbut1} (discounting the rotations of electron spin) there are precisely {\it eight} different types of triplets of order parameters which can be combined into an electrically charged texture. We discuss some examples of these in the concluding section.

Although the matrix $X$, being imaginary in Majorana representation, respects the particle-hole symmetry, it is not the unique such matrix that commutes with the Hamiltonian $H$. There are in fact three more, as dictated by the quaternionic nature of the real representation of the Clifford algebra $C(2,5)$ \cite{remark} furnished by the set of sixteen-dimensional matrices $\{ \Gamma_i, i \Gamma_j \}$, $i=1,2$, $j=0,3,5,6,7$. \cite{herbut1} Therefore, the conservation of the particle-number in the Dirac-Nambu Hamiltonian alone is a necessary but not also a sufficient condition for the finite electrical charge of the texture. The majority of the insulating triplets in graphene will in fact yield precisely zero density of the electrical charge.

It is also interesting to consider the case when the dimension of the matrices in the Hamiltonian $H$ is $d=8$, describing the spinless fermions hopping on honeycomb lattice, or the two boundary surfaces of a thin film of topological insulators, once Nambu's particle-hole doubling is implemented. In this case there are only 10 possible mass terms, and 6 different triplets of mutually anticommuting masses. (Appendix A) Since again $\Gamma_i$  are real for $i=1,2$ and imaginary for $i=0, 3,5$, in this case there are no further matrices that would anticommute with the entire set of $\Gamma_0, \Gamma_i$ $i=1,2,3,5$. \cite{herbut1} The matrix $X$ is then the unique matrix which respects the particle-hole symmetry (that is, which is imaginary in the Majorana representation) and that also commutes with the Dirac-Nambu Hamiltonian. In this case, and in contrast to the case when $d=16$, it is not only necessary but also sufficient that the Hamiltonian $H$ conserves the particle number, in order  for the texture to be electrically charged. In fact, the electrically charged triplet of order parameters is in this case unique.

In the following two sections we prove the Eqs. (4) and (5) for the matrix dimensions of $d=16$ and $d=8$, which involve different types of representations of Clifford algebras and require related, but somewhat different proofs. A reader not interested in the mathematical details may skip over these two sections and go directly to the Section IV. In Appendices A and B we provide some simple (and for our purposes helpful) facts about the requisite Clifford algebras and their relevant representations. In Appendix C we prove our starting assertion that any triplet of anticommuting masses in the sixteen-dimensional Dirac-Nambu Hamiltonian is always a subset of some fivetuplet of compatible (anticommuting) order parameters.

\section{Graphene-like systems: $d=16$}

Let us now prove the Eq. (4). We start with the case relevant to the electrons in graphene, or, more generally, to any other Dirac Hamiltonian deriving from the spin-independent electron hopping on a two-dimensional lattice and respecting parity and time reversal, when the dimension of the matrices is $d=16$.\cite{herbut2} Affleck-Marston $\pi$-flux Hamiltonian on the square lattice,  for example, falls into the same class. The matrices $\Gamma_1, \Gamma_2, i\Gamma_0,  i\Gamma_3, i\Gamma_5, i\Gamma_6, i\Gamma_7$ are then real in Majorana representation, and form an irreducible real representation of the Clifford algebra
$C(2,5)$. \cite{herbut1} The original matrices $\Gamma_1, \Gamma_2, \Gamma_0, \Gamma_3, \Gamma_5, \Gamma_6, \Gamma_7$, however, may alternatively be viewed as forming a particular sixteen-dimensional complex representation of $C(7,0)$. Let us denote this representation by
$Rep_{16} (2,5)$, as a reminder that the representation is 16-dimensional, and that two matrices in it are real, and the remaining five are imaginary. On the other hand, it is well known that  any Clifford algebra $C(2n+1,0)$ has only two inequivalent irreducible complex representations which are both $2^n$-dimensional, and which differ only in the sign of an {\it odd} number of matrices.\cite{okubo} For $C(7,0)$, these will be, in our notation, $Irep_8 (4,3)$ and $Irep'  _8 (4,3)$.  It then follows that either
\begin{equation}
Rep_{16} (2,5) = Irep_8 (4,3) \oplus Irep_8 (4,3),
\end{equation}
or
\begin{equation}
Rep_{16} (2,5) = Irep' _8 (4,3) \oplus Irep' _8 (4,3).
\end{equation}
In other words, the matrices $\Gamma_1, \Gamma_2, \Gamma_0, \Gamma_3, \Gamma_5, \Gamma_6, \Gamma_7$ can be block-diagonalized into an orthogonal sum of two {\it equivalent} irreducible eight-dimensional representations of $C(7,0)$. This is in accord with the fact that $C(2,5)$ has two inequivalent real 16-dimensional representations, which correspond to the two possibilities in Eqs. (7) and (8). This result follows immediately from the observation that the only remaining possibility, namely that
\begin{equation}
Rep_{16} (2,5) = Irep_8 (4,3) \oplus Irep' _8 (4,3),
\end{equation}
i.e. that $\Gamma_\mu = \sigma_3 \otimes \alpha _\mu$, where $\{ \alpha_\mu, \alpha_\nu \} = 2 \delta_{\mu \nu}$ and $\alpha$-matrices form an eight-dimensional irreducible representation of $C(7,0)$, would imply that there are only two matrices which commute with $Rep_{16} (2,5)$: $\sigma_0 \otimes I_8$ and $\sigma_3 \otimes I_8$, where $I_8$ is the eight-dimensional unit matrix. This is because the unit matrix $I_8$ is, by Schur's lemma, the only matrix that commutes with all seven of $\alpha$-matrices. This would contradict, however, the fact that the sixteen-dimensional real representation of $C(2,5)$ is quaternionic, and that consequently $Rep_{16} (2,5)$ actually has {\it four} Casimir operators. \cite{herbut1, okubo, remark2} It is easy to see that both the decompositions in Eqs. (7) and (8) conform to this fact, and these are simply, after the block-diagonalization, $\sigma_0 \otimes I_8$, and $\vec{\sigma} \otimes I_8$.

By a judicious unitary transformation the Dirac-Nambu Hamiltonian $H$ can therefore be transformed into an orthogonal sum of two identical eight-dimensional copies:
\begin{equation}
H= \sigma_0 \otimes i\alpha_0 (\hat{p}_1\alpha_1 + \hat{p}_2 \alpha_2 +i  m_1 + m_2 \alpha_3 + m_3  \alpha_5),
\end{equation}
with the five eight-dimensional $\alpha$-matrices representing $C(5,0)$. But $C(5,0)$ also has only two inequivalent irreducible representations, which are then four dimensional. Again, the very existence of two further anticommuting matrices $\Gamma_6$ and $\Gamma_7$ guarantees that the particular matrices $\alpha_0, \alpha_1, \alpha_2, \alpha_3, \alpha_5$ appearing in the above decomposition must now be an orthogonal sum of two {\it inequivalent} representations of $C(5,0)$. A more detailed proof of the last statement will be given in the Section III when we discuss the case of $d=8$.

In sum, any sixteen-dimensional Dirac-Nambu Hamiltonian $H$, obeying particle-hole symmetry by construction, can be transformed into
\begin{equation}
H=\sigma_0\otimes ( H_0 (\vec{m}) \oplus H_0 (- \vec{m}) )
\end{equation}
where $H_0$ is the irreducible four-dimensional Hamiltonian in Eq. (1).

It then becomes evident that
  \begin{equation}
  \langle \bar{\Psi} \Gamma_\mu (\sigma_0 \otimes \sigma_0\otimes I_4) \Psi \rangle= 0
  \end{equation}
always, because of the cancelation of the two contributions with opposite signs. On the other hand, in the block-diagonal representation in Eq. (11),
 \begin{equation}
  \langle \bar{\Psi} \Gamma_\mu (\sigma_0 \otimes \sigma_3 \otimes I_4) \Psi \rangle= \frac{d}{8} \frac{1}{8\pi} \epsilon_{\mu \nu \lambda} \vec{m} \cdot (\partial_\nu \vec{m} \times \partial_\lambda  \vec{m}),
  \end{equation}
 since the sum of the two contributions now add up. The factor $d/8 = 2$ accounts for the doubling due to spin-1/2 degree of freedom, but not for the particle-hole doubling. The inclusion of the latter would, of course, lead to overcounting. We can also recognize the matrix appearing in the last expression as the matrix $X$ in Eq. (4):
 \begin{widetext}
 \begin{equation}
 X= \sigma_0 \otimes \sigma_3 \otimes I_4 = i(\sigma_0 \otimes \sigma_2 \otimes  i\gamma_1 \gamma_2)  (\sigma_0 \otimes \sigma_1 \otimes  i\gamma_1 \gamma_2),
 \end{equation}
 \end{widetext}
with the two factors $\sigma_2 \otimes  i\gamma_1 \gamma_2$ and $\sigma_1 \otimes  i\gamma_1 \gamma_2$ as precisely the two remaining matrices that anticommute with the set
\begin{equation}
\sigma_0 \otimes  i\gamma_0 \gamma_1, \sigma_0 \otimes  i\gamma_0 \gamma_2, \sigma_3 \otimes \gamma_0, \sigma_3 \otimes  i\gamma_0 \gamma_3, \sigma_3 \otimes  i\gamma_0 \gamma_5,
\end{equation}
which appears in the second factor in Eq. (11) when written explicitly. Reversing the unitary transformations used to bring the Hamiltonian into the final block-diagonal form in Eq. (11),  Eq. (14) implies Eq. (6), or equivalently, Eq. (5).

\section{Topological insulator thin film: $d=8$}

One can similarly establish the validity of Eq. (4)  when the dimension of the matrices in the Hamiltonian is $d=8$,
 with the details of the proof being only slightly  different. The matrices $\Gamma_1, \Gamma_2, i\Gamma_0, i \Gamma_3, i\Gamma_5$ then provide a real eight-dimensional representation of the Clifford algebra $C(2,3)$. (Appendix B)
The original five matrices $\Gamma_1, \Gamma_2, \Gamma_0,  \Gamma_3, \Gamma_5$, on the other hand, can also be viewed as a particular complex representation of $C(5,0)$; call it $Rep_8 (2,3)$. As mentioned earlier, $C(5,0)$ has only two irreducible complex representations; let us call them $Irep_4 (3,2)$ and $Irep' _4 (3,2)$. One can then show that
\begin{equation}
Rep_8 (2,3)= Irep_4 (3,2) \oplus Irep' _4 (3,2),
\end{equation}
i. e. $Rep_8 (2,3)$ can be decomposed into an orthogonal sum of two {\it inequivalent} irreducible representations of $C(5,0)$. Again, this follows from realizing that the remaining possibility that $Rep_8 (2,3)$ is an orthogonal sum of two equivalent representations of $C(5,0)$ would imply that the matrices can be transformed into the form $\Gamma_\mu = \sigma_0 \otimes \gamma_\mu$. This, however, would mean that there is no matrix that anticommutes with all the five matrices in $Rep_8 (2,3)$. This would contradict the fact that $Rep_8 (2,3)$ can be obtained by simply dropping two real matrices in $Ir ep_8 (4,3)$, so that two further matrices which anticommute with the entire representation must exist. (This construction is discussed in Appendix B.) Indeed, from the correct decomposition in Eq. (16) we find these to be $\sigma_1 \otimes I_4$ and $\sigma_2 \otimes I_4$.

The decomposition in Eq. (16) also means that the eight-dimensional real representation of the Clifford algebra $C(2,3)$, up to a unitary transformation of course, is unique.

When $d=8$ we can therefore transform the Dirac-Nambu Hamiltonian into
\begin{equation}
H=H_0 (\vec{m}) \oplus H_0 (- \vec{m}),
\end{equation}
and therefore the induced fermionic current is
\begin{equation}
  \langle \bar{\Psi} \Gamma_\mu (\sigma_3 \otimes I_4) \Psi \rangle= \frac{d}{8} \frac{1}{8\pi} \epsilon_{\mu \nu \lambda} \vec{m} \cdot (\partial_\nu \vec{m} \times \partial_\lambda  \vec{m}).
  \end{equation}
Again, recognizing the sought matrix $X$ in the above as
\begin{widetext}
\begin{equation}
X= \sigma_3 \otimes I_4 =
(\sigma_0\otimes i\gamma_0\gamma_1)
(\sigma_0\otimes i\gamma_0\gamma_2)
(\sigma_3\otimes\gamma_0)
(\sigma_3\otimes i\gamma_0\gamma_3)
(\sigma_3\otimes i\gamma_0\gamma_5),
\end{equation}
\end{widetext}
i. e. as precisely the product of the five matrices which appear in the block-diagonal form of $H$ in Eq. (17), implies the Eq. (5)  in the original representation.

\section{Discussion}

Although formally the expression in Eq. (4) appears the same, the condition for the topological current to be the electrical current when  $d=8$ and when $d=16$ is quite different. One can understand the principal difference in somewhat simpler terms as follows. When $d=8$, the Dirac-Nambu Hamiltonian is formed by the usual Nambu's particle-hole doubling of some original $d=4$ Dirac Hamiltonian which described only particles. The mass-terms in this original four-dimensional representation stand only for the insulating order parameters, and the possible superconducting order parameters are excluded. Since the dimension of matrices is then $d=4$ we can maximally have five mutually anticommuting matrices, and thus there is a unique triplet of insulating masses that can form the textures under consideration. The original result in Eq. (1) then  directly applies. For spinless fermions on honeycomb lattice, the triplet in question consists of two Kekule bond density waves and the charge density wave. For two surfaces of a topological insulator slab, these are the two components of the excitonic insulator \cite{babak} and the surface magnetization. One can also show that when $d=8$ there exist three more superconducting mass order parameters, each contributing two linearly independent mass-matrices. (Appendix A) Finally, there is also  the quantum anomalous Hall state, bringing the total number of possible mass-order parameters to ten. The latter matrix, however, commutes with all of the remaining nine, and thus cannot be used to form the textures we are discussing. The product of all four-dimensional matrices that appear in Eq. (1) is the unit matrix, so Eq. (2) can also be understood as the special case of the Eq. (5), modulo the prefactor in the latter equation.

When $d=16$, the original pre-particle-hole-doubled Dirac Hamiltonian is eight-dimensional, and thus twice the size of the irreducible representation of the five mutually anticommuting matrices, as in Hamiltonian (1). It can be then block-diagonalized either into an orthogonal sum of either two equivalent, or two inequivalent representation of $C(5,0)$. In the latter case, one can always find two matrices which anticommute with the original eight-dimensional Dirac Hamiltonian. Since only the insulating order parameters are discernible in the original $d=8$ representation, this means that the original triplet of insulating orders belongs to the fivetuplet of order parameters which are {\it all} insulating. The product of the five matrices in the Dirac Hamiltonian is then the generalized charge of the texture, which differs from the electrical charge. In the former case, in contrast, the matrices anticommuting with the Dirac Hamiltonian do not exist. That precisely means that upon particle-hole doubling of the $d=8$ Dirac Hamiltonian into the $d=16$ Dirac-Nambu version of itself, two such anticommuting matrices, which are then guaranteed to exist, would have to be superconducting order parameters. The product of the five matrices in the Dirac Hamiltonian is then unity, and the generalized charge is precisely the electrical charge.

 When $d=16$, we can finally view the problem along the lines of the ref. \cite{herbut1} as well. Assuming two mutually anticommuting insulating mass-order-parameters, there are {\it two}  different mutually exclusive ways to complement them to the maximal number of five. \cite{herbut1} There are thus six remaining choices for the third mass that can be combined with the first two into a texture. One fivetuplet to which the two chosen insulators belong is completely insulating. Choosing the third order parameter from that set leads to the non-electrical conserved charge of the texture. The other set has one insulating and two superconducting orders, in addition to the two insulators we began with. Choosing the third matrix to be that single remaining insulator in the mixed fivetuplet endows the texture with the electrical charge.

 Consider, for example, the x- and y-components of the  N\' eel order parameter in graphene. \cite{herbut4}
 If the third order parameter in the texture is the remaining z-component of the same N\' eel order
 parameter, the remaining two orders in the fivetuplet are the two Kekule (singlet) bond-density waves,\cite{ryu, herbut5} and the generalized charge is the ``valley spin". If the third component in the texture is the z-component of the anomalous spin Hall state, on the other hand, the remaining two are the two (phase) components of the z-component of the p-wave superconductor.\cite{honerkamp, roy} The generalized conserved charge is then just the electrical charge.

 If in the previous example the texture is formed out of the two phase components of the z-component of the p-wave superconductor and the z-component of the anomalous spin Hall state, in turn, the generalized charge will be proportional to the z-component of the conventional electron spin, which rotates the x- and y- components of the N\' eel order parameter into each other. Eq. (4) would yield the total value of the generalized charge to be two; since the spin is $\langle S_3 \rangle = \langle \Psi^\dagger X \Psi \rangle/ 2 $, the value of the total integrated third component of the spin correctly comes out as unity.

 \section{Conclusion}

We showed that in graphene-like two-dimensional Dirac systems one can always define a conserved fermionic current induced by a texture in the three anticommuting order parameters, and derived the sufficient condition for the conserved charge to be the electrical charge. Our reasoning was based on the properties of the relevant representations of certain Clifford algebras closed by the matrices appearing in the particle-hole symmetric Dirac-Nambu Hamiltonian with a mass-texture. In particular, from the fact that any three anticommuting order parameters can always be embedded into a larger fivetuplet of such anticommuting order parameters, we derived the expression for the conserved current, Eqs. (4) and (5). The sufficient condition for the electrical charge of the texture is that the two missing order parameters in the fivetuplet are the two phase components of some superconducting order.

 \section{Acknowledgement}

   This work was supported by the NSERC of Canada.

 \section{Appendix A: ten masses for $d=8$}

  We here discuss the masses of the Dirac-Nambu Hamiltonian allowed by the particle-hole symmetry when $d=8$. Consider, the Hamiltonian in the Majorana representation:
  \begin{equation}
  H= \hat{p}_1 R_1  + \hat{p}_2 R_2,
  \end{equation}
 where $R_1$ and $R_2$ are two real, mutually anticommuting, eight-dimensional Hermitian matrices. $H$ is evidently imaginary, and the particle-hole transformation is the operation of complex conjugation.

 We can consider always $R_1$ and $R_2$ as a part of maximally seven mutually anticommuting  eigth-dimensional Hermitian matrices, which we assume all square to unity:
 \begin{equation}
 \{ R_1, R_2, R_3, R_4, I_1, I_2, I_3 \},
 \end{equation}
 where $R_i$ are real, and $I_i$ imaginary.  This is an irreducible complex representation of $C(7,0)$, and $R_4 = i R_1 R_2 R_3 I_1 I_2 I_3$. Given  $R_1$ and $R_2$, one can then form {\it ten} linearly independent imaginary Hermitian matrices that anticommute with both. We organize them as follows:
 \begin{equation}
 \vec{V}_1 = (I_1, I_2, I_3),
 \end{equation}
 \begin{equation}
 \vec{V}_ 2 = (iR_3 I_2 I_3, i R_3 I_1 I_3, i R_3 I_1 I_2  ),
 \end{equation}
 \begin{equation}
 \vec{V}_ 3 = (iR_4 I_2 I_3, i R_4 I_1 I_3, i R_4 I_1 I_2  ),
 \end{equation}
 and
 \begin{equation}
 S= i R_1 R_2.
 \end{equation}
 The last matrix represents the anomalous quantum Hall state which commutes with the others:
 \begin{equation}
 [S, V_{ij}] =0.
 \end{equation}
 Among the remaining  nine,
 \begin{equation}
 \{ V_{mk}, V_{nk} \} = \{ V_{km}, V_{kn} \} = 2 \delta_{mn},
 \end{equation}
 for $k=1,2,3$,  whereas all other pairs commute. There are thus only six possible triplets of  mutually anticommuting mass terms that can form the textures under consideration:
 \begin{equation}
 \sum _{i=1}^ 3 m_i V_{ik},
\end{equation}
or
 \begin{equation}
 \sum _{i=1}^ 3 m_i V_{ki}
\end{equation}
$k=1,2,3$. Out of these six, one and only one will consist of exclusively insulating order parameters, and thus induce the electrical current.

\section{Appendix B: almost complex representation of $C(2,3)$ }

  One can also understand some of the features of the eight-dimensional real representation of $C(2,3)$ used in the text. Namely, from the seven matrices forming the representation of $C(7,0)$ in Eq. (21), we can form
  \begin{equation}
  \{R_1, R_2, iI_1, iI_2, iI_3 \}
  \end{equation}
  as the eight-dimensional real representation of the Clifford algebra $C(2,q)$ with the maximal value of $q$, which is $q=3$. This representation is ``almost complex" in the terminology of Okubo \cite{okubo} since besides the unit matrix it allows one more real matrix that commutes with all the matrices in the representation; it is
  \begin{equation}
  R_3 R_4.
  \end{equation}
   One can then also observe that $Rep_8 (2,3)$, given by Eq. (30) without the factors of the imaginary unit ``i" in the last three terms, necessarily allows for two further anticommuting matrices, as claimed and used in the text: these are $R_3$ and $R_4$.

\section{Appendix C: embedding of a triplet into a fivetuplet for $d=16$}

 The starting point of this work was that given two real sixteen-dimensional anticommuting matrices $\Gamma_1$ and $\Gamma_2$, and three anticommuting imaginary matrices $\Gamma_0$, $\Gamma_3$, and $\Gamma_5$, one can always find two further imaginary matrices $\Gamma_6$ and $\Gamma_7$, so that any chosen pair out of the seven matrices anticommutes.

   The proof follows from observing that the set $\{ \Gamma_i, i \Gamma_j \}$, $i=1,2$ and $j=0,3,5$ represents a sixteen-dimensional real reepresentation of the Clifford algebra $C(2,3)$. $C(2,3)$, on the other hand, has a unique real irreducible representation which is eight dimensional. \cite{okubo} We saw this already as following from the decomposition in Eq. (16), for example. Therefore, by a unitary transformation we can write
   \begin{equation}
   \Gamma_i = \sigma_0 \otimes \alpha_i
   \end{equation}
   where $\alpha_i$ are eight-dimensional, and real for $i=1,2$, and imaginary for $i=0,3,5$. Since in the eight-dimensional space there exist then two additional linearly independent real matrices, call them $\alpha_6$ and $\alpha_7$ (Appendix B), two additional imaginary anticommuting matrices are,
   \begin{equation}
   \Gamma_6= \sigma_2 \otimes \alpha_6,
   \end{equation}
   and
  \begin{equation}
   \Gamma_7 = \sigma_2 \otimes \alpha_7.
   \end{equation}
 The pair is obviously unique. Three non-trivial Casimir operators for the real representation of $C(2,5)$ given by $\{ \Gamma_i, i \Gamma_j \}$, $i=1,2$ and $j=0,3,5,6,7 $ are then $\sigma_1 \otimes \alpha_6 \alpha_7$, $\sigma_3 \otimes \alpha_6 \alpha_7$, and $i\sigma_2 \otimes I_8$. Together with the unit matrix  they close the quaternionic algebra.


\begin{thebibliography}{99}
\bibitem{remark0} The large number of 36 can be understood as 16 insulating + 20 superconducting linearly independent masses; there are only four types of insulators, each being possible in the singlet and the triplet versions ($16=4
\times  4$), and four different types of superconductors, one singlet (contributing $2$ mass terms for x- ant y- components of the order parameter)
and three triplets  (contributing further $3\times 2 \times  3 = 18$ mass terms).
\bibitem{remark} $C(p,q)$ is defined as a set of $p+q$ mutually anticommuting
 generators, $p$ ($q$) of which square to $+1$ ($-1$).
\bibitem{remark2} It is possible to have non-trivial Casimir operators for the representation in question because
the irreducibility is defined here with respect to other purely real, and not complex (as in Eq. (7)), smaller representations.
See also Appendix C for an example.
\end{thebibliography}
\end{document}